\documentstyle[amssymb,prl,aps]{revtex}

\begin{document}
\title{Coherent and incoherent pumping of electrons in double quantum dots}
\author{B.L. Hazelzet, M.R. Wegewijs, T.H. Stoof and Yu. V. Nazarov}
\address{Department of Applied Physics, Faculty of Applied Science,\\
Delft University of Technology, Lorentzweg 1, 2628 CJ Delft,The Netherlands}
\maketitle

\begin{abstract}
We propose a new mode of operation of an electron pump consisting of two
weakly coupled quantum dots connected to reservoirs. An electron can be
transferred within the device at zero bias voltage when it is subjected to
electromagnetic radiation, thereby exciting the double dot. The excited
state can decay by transferring charge from one lead and to the other lead
in one direction. Depending on the energies of the intermediate states in
the pumping cycle, which are controlled by the gate voltages, this transport
is either incoherent via well-known sequential tunneling processes, or
coherent via a inelastic co-tunneling process. The latter novel mode of
operation is possible only when interdot Coulomb charging is important. The
D.C. transport through the system can be controlled by the frequency of the
applied radiation. We concentrate on the resonant case, when the frequency
matches the energy difference for exciting an electron from one dot into the
other. The resonant peaks in the pumping current should be experimentally
observable. We have developed a density matrix approach which describes the
dynamics of the system on timescales much larger than the period of the
applied irradiation. In contrast to previous works we additionally consider
the case of slow modulation of the irradiation amplitude. Harmonic
modulation produces additional sidepeaks in the photoresponse, and pulsed
modulation can be used to resolve the Rabi frequency in the time-averaged
current.
\end{abstract}

\section{Introduction}

In recent years quantum dots have attracted great attention. A quantum dot
can be thought of as an artificial atom with adjustable parameters. It is of
more than fundamental interest to study its properties under various
circumstances, e.g. by transport experiments~\cite{bib:leorev}. By
considering a double-quantum-dot system, the analogy with real atoms can be
stretched to include artificial molecules. The analogue of the covalent bond
is then formed by an electron which coherently tunnels back and forth
between the two dots. By applying electromagnetic radiation with a frequency
equal to the level detuning in the double-dot system, an electron can
undergo so-called spatial Rabi oscillations even when the tunneling matrix
element between the dots is small.

Recently, several time-dependent transport measurements on quantum-dot
systems have been reported~\cite{bib:spectro,bib:Oosterkamp}, most of them
being of a spectroscopic nature. It has also been suggested to construct
devices from quantum dots. Examples of such applications are pumps that
transfer electrons one by one at zero bias voltage by using time-dependent
voltages to raise and lower tunnel barrier heights~\cite{bib:leo40}, or
systems in which coupled quantum dots (or quantum wells) are used for
quantum-scale information processing~\cite{bib:qcomp}. Several theoretical
models for time-dependent transport through a double quantum dot have
already been proposed. For instance, in Ref.~\cite{bib:StaffordWingreen} D.
C. transport was considered for arbitrary bias voltage when the signal
couples to the gate voltages of the dots. At zero bias voltage the system
operates as an electron pump. In \ Ref.~\cite{bib:StoofNazarov} the D. C.
current, controlled by external irradiation, was considered for finite bias
voltage. These results were recently generalized to include time-dependent
gate and bias voltages and tunnel barriers~\cite
{bib:BruneBruderSchoeller,bib:ZieglerBruderSchoeller}. In all these works a
tunneling-Hamiltonian approach was used to incorporate the effects of
Coulomb interaction between electrons on the same and on different dots. It
is assumed that the barriers separating the leads and the dots are high and
therefore the wave functions have only a small overlap. A different approach
would be to use scattering states of electrons which extend through the
leads and dots which is appropriate for almost transparent barriers (see
e.g.~\cite{bib:PedersenButtiker}). However, the effects of Coulomb
interaction are not easily taken into account in this approach. Because the
transport mechanism in the above mentioned double-dot pumps is determined by
sequential tunneling, electrons are pumped {\em incoherently} i.e. the
tunneling of an electron into and out of the device are independent events.
Also, in these works the time-dependent signal is taken to be a
monochromatic.

In this paper we describe,\ firstly, an new mode of operation of such an
electron pump. In this case, electrons can be transferred coherently through
the system by means of a co-tunneling process. Our device has to be designed
in such a way that interdot Coulomb repulsion is important. By appropriately
adjusting the gate voltages the device can be made to operate in the
co-tunneling regime or the sequential tunneling regime. The latter regime is
considered here for comparison with previous works and should be
distinguished from the coherent one. In the co-tunneling regime the interdot 
{\em attraction }between an extra electron excited by the A. C. field into
one dot and the hole it left in the other dot stabilizes the excited state.
Electrons are prevented from entering or leaving the device independently
and inelastic co-tunneling is the lowest non-vanishing order process for
transport. This involves correlated tunneling events through the different
junctions connecting the dots weakly to the leads. The system switches {\em %
coherently} from the excited state to the ground state each time an electron
is pumped through the dots. Alternatively, this process can be seen as
coherent transport through a double-quantum-dot qubit.\ The second main
result of this paper is that we predict the effect of slow modulation of the
irradiation amplitude on the pumping current. The timescale of the amplitude
modulation is assumed to be much larger than the one set by the frequency of
the unmodulated signal.

The paper is organized as follows: in Sec. \ref{sec:system} we introduce the
system. In Sec. \ref{sec:timescale} we develop the density matrix approach
to describe the system only on a timescale much larger than the period of
the applied irradiation. This development is similar to that in
quantum-optics for resonance fluorescence e.g.~\cite{bib:qopt} and is
central to our treatment of the problems. We apply this approach to the
transport through the double-dot in the sequential and the co-tunneling
regime. In Sec. \ref{sec:modulation} we consider the case where the
amplitude of the applied radiation is modulated on the large timescale.
Finally, a summary and conclusions are presented in Sec. \ref{sec:conclusion}%
.

\section{Double-dot electron pump}

\label{sec:system}The system we consider consists of two coherently coupled
quantum dots $1$\ and $2$\ connected by tunnel barriers to large reservoirs $%
L$ and $R,$\ as depicted in Fig. \ref{fig:2dotspump}. The leads are assumed
to have a continuous electronic energy spectrum. The fixed difference
between the electro-chemical potentials of the leads $\mu _{L}=\mu
-eV_{L_{1}}$\ and $\mu _{R}=\mu -eV_{L_{2}}$\ ($e>0$) and the temperature
are the smallest two energy scales of the problem. We can thus take them to
be zero. Each quantum dot $i=1,2$\ contains some number $N_{i}$\ electrons
and is assumed to be in the ground state. We will concentrate on transitions
between the ground states of the individual dots with different numbers of
electrons.

Disregarding all tunneling for the moment, let us consider the energy of the
double dot within\ the standard Coulomb blockade model. The capacitive
coupling between dot $i=1,2$\ and the attached electrodes is taken into
account by the gate capacitance $c_{G_{i}}$and the lead capacitance $%
c_{L_{i}}$. The mutual capacitive coupling between the dots is described by
the interdot capacitance $c_{12}$. The total energy of the double dot
system, when dot $1$\ and dot $2$\ are respectively in the $N_{1}$\ and $%
N_{2}$\ electron ground state, reads (see e.g. Ref. \cite
{bib:ZieglerBruderSchoeller}): 
\begin{equation}
E_{N_{1},N_{2}}=\sum_{i=1,2}\sum_{l=1}^{N_{i}}\varepsilon _{il}+\sum_{i=1,2}%
\frac{1}{2}u_{ii}N_{i}\left( N_{i}-1\right) +u_{12}N_{1}N_{2}
\label{eq:E[N[1],N[2]]}
\end{equation}
In the first term, $\varepsilon _{il}$\ is the $l$th effective single
particle energy, $l=1,\ldots ,N_{i}$\ in dot $i=1,2$: 
\[
\varepsilon _{il}\left( V_{G_{1}},V_{G_{2}}\right) =\varepsilon _{il}^{0}+%
\frac{1}{2}u_{ii}-e\sum_{j=1,2}\left( \alpha _{iG_{j}}V_{G_{j}}+\alpha
_{iL_{j}}V_{L_{j}}\right) 
\]
This incorporates the bare single particle energy $\varepsilon _{il}^{0}$, a
conveniently chosen offset $u_{ii}/2$\ and the linear shift with the
electrode voltages. The coefficients of the voltages $\alpha
_{iG_{j}}=C_{ij}^{-1}c_{G_{j}},\alpha _{iL_{j}}=C_{ij}^{-1}c_{L_{j}}$\
depend on the dot-electrode capacitances and the inverse capacitance matrix
elements 
\[
C_{ii}^{-1}=\frac{\left( c_{1}c_{2}\right) /c_{i}}{c_{1}c_{2}-c_{12}^{2}}%
,\qquad C_{12}^{-1}=C_{21}^{-1}=\frac{c_{12}}{c_{1}c_{2}-c_{12}^{2}},
\]
where $c_{i}=c_{G_{i}}+c_{L_{i}}+c_{12}$ is the total capacitance of dot $%
i=1,2$.\ By appropriately changing the gate voltages the effective single
particle levels in dot $1$\ and $2$\ can be independently be shifted with
respect to each other. In the second and third term in equation (\ref
{eq:E[N[1],N[2]]}), $u_{ii}=e^{2}C_{ii}^{-1}$\ is the {\em intra}dot
charging energy of dot $i=1,2$\ and $u_{12}=e^{2}C_{12}^{-1}<u_{11},u_{22}$\
the {\em inter}dot charging energy.

Let us now consider the stability of a ground state of the double dot with
respect to the tunneling between the individual dots and the leads. Assume
that the gate voltages $V_{G_{1}},V_{G_{2}}$\ are such that in the stable
state of the double dot there are $N_{1}$\ and $N_{2}$\ electrons in,
respectively, dot $1$\ and $2$\ and denote this state by $|0,0\rangle $. The
stability diagram of the double dot near the region were this state is
stable is sketched in Fig. \ref{fig:stab}\ for the typical case where
interdot charging is important: $u_{12}\lesssim u_{11},u_{22}$.\ The region
of stability for state $|0,0\rangle $\ has a hexagonal shape. The stable
states $|n_{1},n_{2}\rangle $ in the six neighboring regions have $%
N_{1}+n_{1}$\ and $N_{2}+n_{2}$\ electrons in dot $1$\ and $2$,
respectively, where $n_{1}\neq n_{2}=0,\pm 1$\ and $\left|
n_{1}+n_{2}\right| \leq 1$. The energies of these states, $%
E_{n_{1},n_{2}},n_{i}=0,1$,\ are found from the right-hand-side of (\ref
{eq:E[N[1],N[2]]}) by replacing $N_{i}\rightarrow N_{i}+n_{i}$. The
hexagonal region is bounded by six stability constraints for state $%
|0,0\rangle $. The first four constraints follow from the requirement that
the energy barrier for injection ($+$) or emission ($-$) of an electron to
or from either lead,%
\begin{mathletters}%
%
\begin{eqnarray*}
\Delta _{L}^{+} &=&E_{1,0}-\mu -E_{0,0}=\varepsilon
_{1N_{1}+1}+u_{11}N_{1}+u_{12}N_{2}-\mu , \\
\Delta _{R}^{+} &=&E_{0,1}-\mu -E_{0,0}=\varepsilon
_{2N_{2}+1}+u_{22}N_{2}+u_{12}N_{1}-\mu , \\
\Delta _{R}^{-} &=&E_{0,-1}+\mu -E_{0,0}=\mu -\left( \varepsilon
_{2N_{2}}+u_{22}\left( N_{2}-1\right) +u_{12}N_{1}\right) , \\
\Delta _{L}^{-} &=&E_{-1,0}+\mu -E_{0,0}=\mu -\left( \varepsilon
_{1N_{1}}+u_{11}\left( N_{1}-1\right) +u_{12}N_{2}\right) ,
\end{eqnarray*}
\label{eq:barriers_seq}%
\end{mathletters}%
%
should be positive. Sufficiently far away from the four lines $\Delta
_{L,R}^{\pm }=0$\ i.e. $\Delta _{L,R}^{\pm }\gg \Gamma _{L,R}$\ the
sequential tunneling of single electrons through the junctions connecting a
dot and lead is suppressed, cf. Fig. \ref{fig:stab}. Here the typical tunnel
rate is $\Gamma _{L,R}=2\pi v_{L,R}\left| t_{L,R}\right| ^{2}$\ where $%
v_{L,R}$\ is the density of states in the left and right lead respectively
and $t_{L,R}$\ is the matrix element between the states in the lead and in
the dot, which depends only weakly on the energy. Two of sequential
tunneling barriers (\ref{eq:barriers_seq}) can be independently tuned by the
gate voltages, the other two are related to these by $\Delta _{L}^{+}+\Delta
_{L}^{-}=u_{11}+\delta _{1}$\ and $\Delta _{R}^{+}+\Delta
_{R}^{-}=u_{22}+\delta _{2}$\ where\ $\delta _{i}=\varepsilon
_{iN_{i}+1}-\varepsilon _{iN_{i}}=\varepsilon _{iN_{i}+1}^{0}-\varepsilon
_{iN_{i}}^{0}\ $is the spacing between the two ``active'' single particle
levels in dot $i=1,2$. It will be convenient from here on to consider $%
0<\Delta _{L}^{-}<u_{11}+\delta _{1}$\ and $0<\Delta _{R}^{+}<u_{22}+\delta
_{2}$\ as independent variables instead of the two gate voltages, cf. Fig. 
\ref{fig:stab}. Two additional stability constraints follow from the
requirement that the energy barriers for polarizing the double dot with
respect to state $|0,0\rangle $,%
\begin{mathletters}%
%
\begin{eqnarray}
\varepsilon _{0} &=&E_{-1,1}-E_{0,0}=\Delta _{L}^{-}+\Delta _{R}^{+}-u_{12},
\label{eq:epsilon[0]} \\
\varepsilon _{0}^{\prime } &=&E_{1,-1}-E_{0,0}=\Delta _{L}^{+}+\Delta
_{R}^{-}-u_{12},  \label{eq:epsilon[0]'}
\end{eqnarray}
\label{eq:barriers_cot}%
\end{mathletters}%
%
should be positive. Here $\Delta _{L}^{\mp },\Delta _{R}^{\pm }$\ are the
positive Coulomb energies we must pay to change the number of electrons on
each dot from the stable configuration $|0,0\rangle $,\ and $-u_{12}$\ is
the energy we gain by creating an attracting electron-hole pair with respect
to the stable state $|0,0\rangle $. Note that the former energy also depends
on $u_{12}$\ i.e. it incorporates the interaction between the extra electron
or hole and the electrons in{\em \ both} dots. Sufficiently far from the
lines $\varepsilon _{0}=0$\ and $\varepsilon _{0}^{\prime }=0$, cf.\ Fig. 
\ref{fig:stab}, the polarization of the double dot by a coherent
co-tunneling process is suppressed: $\varepsilon _{0},\varepsilon
_{0}^{\prime }\gg \Gamma _{\text{ct}}^{\prime }$ where $\Gamma _{\text{ct}%
}^{\prime }\ll \Gamma _{L,R}$ is some typical rate for this process. In such
a second order process an electron is injected into one dot and another
electron is emitted from the other dot, effectively transporting one charge
across the double dot. From the relation $\varepsilon _{0}+\varepsilon
_{0}^{\prime }=\sum_{i=1,2}\left( u_{ii}-u_{12}+\delta _{i}\right) >0$\ we
find that $\varepsilon _{0},\varepsilon _{0}^{\prime }>0$\ corresponds to $%
u_{12}<\Delta _{L}^{-}+\Delta _{R}^{+}<\sum_{i=1,2}\left(
u_{ii}-u_{12}+\delta _{i}\right) +u_{12}$ in the stability diagram.

Now consider the coherent tunneling of electrons between the dot $1$\ and $2$%
. If the co-tunneling barriers are also larger than the matrix element $T$
for this process, $\varepsilon _{0},\varepsilon _{0}^{\prime }\gg T,\ $then
the polarization of the double dot by coherent tunneling of an electron
between the dots is also suppressed. Under these conditions the D.C.
transport through the double dot is blocked at low bias voltage i.e. we have
the Coulomb blockade.

This situation is changed, however, if we apply electromagnetic radiation to
the system. Assume that a time-dependent oscillating signal is present on
the gate electrodes which will shift the energies of single levels without
altering the wave functions too much, so that the time-dependent energy
difference between states $|0,0\rangle $\ and $|-1,1\rangle $\ becomes 
\begin{equation}
\varepsilon (t)=\varepsilon _{0}+V\cos \omega t,
\end{equation}
where $V$\ is the amplitude and $\omega $\ the frequency of the externally
applied signal. When the frequency of this applied radiation matches the
time-independent energy difference $\varepsilon _{0}$\ between states $%
|0,0\rangle $\ and $|-1,1\rangle $, it is possible for an electron from the
left dot to tunnel to the right one by absorbing one energy quantum $\omega
\approx \varepsilon _{0}$\ from the field. In principle, this electron can
now leave the system by tunneling to the right lead, resulting in the state $%
|-1,0\rangle $. An electron from the left lead can then tunnel to the left
dot, thus restoring the ground state. Effectively, an electron has now been
transferred from the left electrode to the right one. Alternatively, the
electron can coherently tunnel back by emitting an energy quantum resulting
in state $|0,0\rangle $. This transport cycle, $|0,0\rangle \leftrightarrow
|-1,1\rangle \rightarrow |-1,0\rangle \rightarrow |0,0\rangle $, is not the
only one. Another possible sequence, in which the system passes the
intermediate state $|0,1\rangle $, is given by $|0,0\rangle \leftrightarrow
|-1,1\rangle \rightarrow |0,1\rangle \rightarrow |0,0\rangle $. The tree
other states can be disregarded under the following conditions. Firstly, the
field should not be resonant with the transition to the other excited state $%
|1,-1\rangle $:\ $\left| \varepsilon _{0}-\omega \right| \ll \left|
\varepsilon _{0}^{\prime }-\omega \right| $.$\ $This is the case when the
distance between the two excited levels is much larger than the frequency
detuning of the frequency, $\left| \varepsilon _{0}^{\prime }-\varepsilon
_{0}\right| \gg \left| \delta \omega \right| =\left| \varepsilon _{0}-\omega
\right| $, which, in the stability diagram, corresponds to (cf. equations (%
\ref{eq:epsilon[0]}), (\ref{eq:epsilon[0]'})) 
\begin{equation}
\left| 2\left( \Delta _{L}^{-}+\Delta _{R}^{+}\right) -\sum_{i=1,2}\left(
u_{ii}+\delta _{i}\right) \right| \gg \left| \delta \omega \right| .
\label{eq:cond_excite}
\end{equation}
Secondly, the states $|1,0\rangle ,|0,-1\rangle $\ should not be resonant
with intermediate states of the transport cycle $|0,1\rangle $\ and $%
|-1,0\rangle $, respectively: $\left| E_{1,0}-E_{0,1}\right| \gg T$\ and{\bf %
\ }$\left| E_{0,-1}-E_{-1,0}\right| \gg T$ corresponding to%
\begin{mathletters}%
%
\begin{eqnarray}
\left| u_{11}+\delta _{1}-\left( \Delta _{L}^{-}+\Delta _{R}^{+}\right)
\right| &\gg &T, \\
\left| u_{22}+\delta _{2}-\left( \Delta _{L}^{-}+\Delta _{R}^{+}\right)
\right| &\gg &T.
\end{eqnarray}
\label{eq:cond_extra}%
\end{mathletters}%
%
Thus under these conditions an electron can be excited from the left dot
into the right one, but the probability of exciting an electron from the
right dot to the left dot can be disregarded.

The details of the transport mechanism of a pumping cycle depend on the
energies of the intermediate states $|-1,0\rangle $\ and $|0,1\rangle $\
relative to the pumped state $|-1,1\rangle $ as shown in Fig. \ref
{fig:energies} which are controlled by the gate voltages.\ The energy
barrier for injecting an electron from the left lead into dot $1$\ and for
emitting an electron from the dot $2$\ dot to the right lead,%
\begin{mathletters}%
%
\begin{eqnarray}
\tilde{\Delta}_{L}^{+} &=&E_{0,1}-\mu -E_{-1,1}=u_{12}-\Delta _{L}^{-}, \\
\tilde{\Delta}_{R}^{-} &=&E_{-1,0}+\mu -E_{-1,1}=u_{12}-\Delta _{R}^{+},
\end{eqnarray}
\label{eq:barriers_cot_excited}%
\end{mathletters}%
%
can be positive or negative, depending on the position in the stable region
of $|0,0\rangle $. Here $u_{12}$\ is the energy we must pay to break up the
attracting electron-hole pair with respect to state $|0,0\rangle $ and $%
-\Delta _{L}^{-},-\Delta _{R}^{+}$\ is the Coulomb energy we gain by
changing the number of electrons on one of the dots to the value of the
stable configuration $|0,0\rangle $. In this paper we consider two regimes
of operation of the double dot electron pump, which are schematically
depicted in Fig. \ref{fig:energies}: (I) both barriers are negative, $\tilde{%
\Delta}_{L}^{+},\tilde{\Delta}_{R}^{-}\ll -\Gamma _{L},-\Gamma _{R}$:\ the
pumped level can decay through sequential tunneling processes; (II) both
barriers are positive, $\tilde{\Delta}_{L}^{+},\tilde{\Delta}_{R}^{-}\gg
\Gamma _{L},\Gamma _{R}$: the excited state is stable with respect to
sequential tunneling but can decay through an inelastic co-tunneling
process. We do not consider the more complicated intermediate case $\left| 
\tilde{\Delta}_{L}^{+}\right| ,\left| \tilde{\Delta}_{R}^{-}\right| \lesssim
\Gamma _{L},\Gamma _{R}$ where resonant processes between leads and dots are
important. In the stability diagram in Fig. \ref{fig:stab} the two regimes
correspond to%
\begin{mathletters}%
%
\begin{eqnarray}
\text{(I)}\quad \left| \Delta _{L}^{-}-\Delta _{R}^{+}\right| &<&\left(
\Delta _{L}^{-}+\Delta _{R}^{+}\right) -2u_{12},  \nonumber \\
2u_{12} &<&\Delta _{L}^{-}+\Delta _{R}^{+}<\sum_{i=1,2}\left(
u_{ii}-u_{12}+\delta _{i}\right) ,  \label{eq:regimeI} \\
\text{(II)}\quad \left| \Delta _{L}^{-}-\Delta _{R}^{+}\right|
&<&2u_{12}-\left( \Delta _{L}^{-}+\Delta _{R}^{+}\right) ,  \nonumber \\
u_{12} &<&\Delta _{L}^{-}+\Delta _{R}^{+}<2u_{12},  \label{eq:regimeII}
\end{eqnarray}
\end{mathletters}%
%
where $<$\ stands for ``separated by energy large compared with $\Gamma
_{L,R}$''. The narrow strips defined by conditions (\ref{eq:cond_excite})
and (\ref{eq:cond_extra}) should be excluded from these regions.

In regime I ($\tilde{\Delta}_{L}^{+},\tilde{\Delta}_{R}^{-}<0$) the charging
of the individual dots dominates the transport. The system relaxes to the
ground state via the two sequential (and thus incoherent) tunneling
processes described above. As shown in Fig. \ref{fig:energies}(a), the four
tunneling processes are described by the rates $\Gamma _{L}^{1},\Gamma
_{R}^{1},\Gamma _{L}^{0}$, and $\Gamma _{R}^{0}$, respectively. The
following rate equations describe the density matrix in the sequential
tunneling regime:%
\begin{mathletters}%
%
\begin{eqnarray}
\partial _{t}\rho _{-10,-10} &=&+\Gamma _{R}^{0}\rho _{-11,-11}-\Gamma
_{L}^{0}\rho _{-10,-10} \\
\partial _{t}\rho _{00,00} &=&-\text{i}T(\rho _{-11,00}-\rho
_{00,-11})+\Gamma _{L}^{0}\rho _{-10,-10}+\Gamma _{R}^{1}\rho _{01,01}, \\
\partial _{t}\rho _{-11,-11} &=&+\text{i}T(\rho _{-11,00}-\rho
_{00,-11})-\left( \Gamma _{R}^{0}+\Gamma _{L}^{1}\right) \rho _{-11,-11}, \\
\partial _{t}\rho _{01,01} &=&+\Gamma _{L}^{1}\rho _{-11,-11}-\Gamma
_{R}^{1}\rho _{01,01}, \\
\partial _{t}\rho _{-11,00} &=&-\text{i}T(\rho _{00,00}-\rho _{-11,-11})-%
\text{i}\varepsilon \left( t\right) \rho _{-11,00}-%
{\textstyle{1 \over 2}}%
\left( \Gamma _{L}^{1}+\Gamma _{R}^{0}\right) \rho _{-11,00},
\end{eqnarray}
\label{eq:seq}%
\end{mathletters}%
%
Here, and throughout this paper, units are used such that $\hbar =1$. The
diagonal elements give the probabilities for an electron to be in the
corresponding states and probability is conserved i.e. at any time $t$ 
\begin{equation}
\rho _{-10,-10}+\rho _{00,00}+\rho _{-11,-11}+\rho _{01,01}=1\text{.}
\end{equation}
The non-diagonal elements$\ \rho _{-11,00}=\rho _{00,-11}^{\ast }$ describe
the coherence between states $|-1,1\rangle $ and $|0,0\rangle $. In general
the tunnel rates $\Gamma _{L}^{0,1}$ and $\Gamma _{R}^{0,1}$ depend on the
energy difference between the states of the transition. We can take $\Gamma
_{L,R}^{0,1}=\Gamma _{L,R}=2\pi v_{L,R}\left| t_{L,R}\right| ^{2}$\ when the
density of states $v_{L,R}$\ in the left and right lead respectively and the
matrix element $t_{L,R}$\ between the states in the lead and in the dots
depends only weakly on the energy. The average current through the system is
given by 
\begin{equation}
I/e=\Gamma _{R}^{0}\rho _{-11,-11}+\Gamma _{R}^{1}\rho _{01,01}\text{.}
\label{eq:current}
\end{equation}

In regime II ($\tilde{\Delta}_{L}^{+},\tilde{\Delta}_{R}^{-}>0$) the
interdot attraction of the electron-hole pair is dominant over the
individual (de)charging of the individual dots. The decay of the excited
state $|-1,1\rangle $ via sequential tunneling is blocked as shown in Fig. 
\ref{fig:energies}(b). However, transport is still possible via inelastic
co-tunneling of electrons~\cite{bib:cotunnel}. When the system is in state $%
|-1,1\rangle $, two electrons can tunnel simultaneously through different
barriers, one going from the left lead to the left dot, and one from the
right dot to the right lead. Because in this transport process a state is
virtually occupied these two tunneling events cannot be treated
independently. The necessary energy is provided by relaxing the system to
the ground state $|0,0\rangle $, thereby releasing an energy $E\approx
\varepsilon _{0}.$The co-tunneling rate can be calculated with Fermi's
Golden Rule. The relevant matrix element is a sum of matrix elements for the
two possible coherent processes which transfer one electron from $L$ to $R$.
The co-tunnel rate is obtained by integrating the partial rates for
transitions over the different final states of the leads which are assumed
to be uncorrelated: 
\begin{eqnarray}
\Gamma _{\text{ct}} &=&2\pi \nu _{L}\nu _{R}\int_{-\infty }^{\infty
}d\varepsilon _{L}\int_{-\infty }^{\infty }d\varepsilon _{R}f\left(
\varepsilon _{L}\right) \left( 1-f\left( \varepsilon _{R}\right) \right) \\
&&\times \left| \frac{t_{L}t_{R}}{\tilde{\Delta}_{L}^{+}-\varepsilon _{L}}+%
\frac{t_{L}t_{R}}{\tilde{\Delta}_{R}^{-}+\varepsilon _{R}}\right| ^{2}\delta
\left( \varepsilon _{0}+\varepsilon _{L}-\varepsilon _{R}\right)
\end{eqnarray}
Here the matrix elements $t_{\text{L,R}}$ for tunneling through the left and
right barrier and the densities of states $\nu _{L,R}$ in the left and right
electrode are assumed to be energy independent. The zero-temperature
co-tunnel rate in our electron pump is%
\begin{mathletters}%
%
\begin{eqnarray}
\Gamma _{\text{ct}} &=&\frac{\Gamma _{L}\Gamma _{R}}{2\pi }\left[ \frac{%
\varepsilon _{0}}{\left( \tilde{\Delta}_{L}^{+}+\varepsilon _{0}\right) 
\tilde{\Delta}_{L}^{+}}+\frac{\varepsilon _{0}}{\tilde{\Delta}_{R}^{-}\left( 
\tilde{\Delta}_{R}^{-}+\varepsilon _{0}\right) }\right.  \nonumber \\
&&\left. +2\frac{\ln \left( 1+\frac{\varepsilon _{0}}{\tilde{\Delta}_{L}^{+}}%
\right) +\ln \left( 1+\frac{\varepsilon _{0}}{\tilde{\Delta}_{R}^{-}}\right) 
}{\tilde{\Delta}_{L}^{+}+\tilde{\Delta}_{R}^{-}+\varepsilon _{0}}\right]
\label{eq:gamma[ct]} \\
&=&\frac{\Gamma _{L}\Gamma _{R}}{2\pi }\varepsilon _{0}\left( \frac{1}{%
\tilde{\Delta}_{L}^{+}}+\frac{1}{\tilde{\Delta}_{R}^{-}}\right) ^{2}+O\left(
\varepsilon _{0}^{2}\right)
\end{eqnarray}
\end{mathletters}%
%
Note that from $\varepsilon _{0}=u_{12}-\left( \tilde{\Delta}_{L}^{+}+\tilde{%
\Delta}_{R}^{-}\right) $\ (cf. equations (\ref{eq:epsilon[0]}) and (\ref
{eq:barriers_cot_excited})) we observe that within regime II we can have $%
\varepsilon _{0}\sim \tilde{\Delta}_{L}^{+},\tilde{\Delta}_{R}^{-}$. The
energy denominators in (\ref{eq:gamma[ct]}) reflect the fact that the
tunneling occurs via the virtual occupation of two states. In contrast to
the incoherent sequential tunneling mechanism, the only relevant density
matrix elements are those between states $|0,0\rangle $ and $|-1,1\rangle $
since states $|-1,0\rangle $ and $|0,1\rangle $ are occupied only virtually.
Taking the co-tunneling processes into account we obtain the equations of
motion for the density matrix elements:%
\begin{mathletters}%
%
\begin{eqnarray}
\partial _{t}\rho _{00,00} &=&-\text{i}T(\rho _{-11,00}-\rho
_{00,-11})+\Gamma _{\text{ct}}\rho _{-11,-11} \\
\partial _{t}\rho _{-11,-11} &=&+\text{i}T(\rho _{-11,00}-\rho
_{00,-11})-\Gamma _{\text{ct}}\rho _{-11,-11} \\
\partial _{t}\rho _{-11,00} &=&-\text{i}T(\rho _{00,00}-\rho _{-11,-11})-%
\text{i}\varepsilon \left( t\right) \rho _{-11,00}-%
{\textstyle{1 \over 2}}%
\Gamma _{\text{ct}}\rho _{00,-11},
\end{eqnarray}
\label{eq:cot}%
\end{mathletters}%
%
where $\rho _{-11,00}=\rho _{00,-11}^{\ast }$ and the probability is
conserved: 
\[
\rho _{00,00}+\rho _{-11,-11}=1 
\]
The current is 
\begin{equation}
I\left( t\right) /e=\Gamma _{\text{ct}}\rho _{-11,-11}\left( t\right) \text{.%
}  \label{eq:endcur}
\end{equation}

In both regimes the irradiation relaxes the constraint of energy
conservation during tunneling by allowing an electron to absorb or emit a
multiple of the energy quantum $\omega $. This gives rise to an enhancement
of the zero bias D.C. component of the current and additionally it
introduces fast oscillations with small amplitude which do not interest us.
In the next section we show how to extract the slowly varying component of
the density matrix from the exact equations (\ref{eq:seq})\ and (\ref{eq:cot}%
) respectively. We point out that above we have written expressions for the
particle current only. The displacement current can be disregarded here
since it does not contribute to the D. C. current.

\section{Timescale separation}

\label{sec:timescale}In this section we consider the dynamics of the two
coherently coupled levels $|a\rangle =|0,0\rangle ,|b\rangle =|-1,1\rangle $
on timescales much larger than the period of the applied irradiation $%
t_{\omega }=2\pi /\omega $. The details of the other states in each regime
are only important for the incoherent processes which depend only weakly on
the time-dependent energy difference between basis states $\left|
a\right\rangle $ and $\left| b\right\rangle $%
\begin{equation}
\varepsilon \left( t\right) =\varepsilon _{0}+V(t)\cos \left( \omega t\right)
\end{equation}
Here we also allow the amplitude $V(t)$ to be modulated on a timescale which
is large relative to $t_{\omega }$. The timescale separation for both
regimes can be done in the same way. The coherent part of the dynamics of
the state of the system depends only on the\ time-dependent: $H\left(
t\right) =H_{0}\left( t\right) +H_{T}$\ where 
\begin{equation}
{\cal H}_{0}\left( t\right) =\frac{1}{2}\varepsilon \left( t\right) \left(
\left| a\right\rangle \left\langle a\right| -\left| b\right\rangle
\left\langle b\right| \right)
\end{equation}
introduces the energy difference between double dot states with zero extra
electrons and $H_{T}$ describes the tunnel coupling between the dots: 
\begin{equation}
{\cal H}_{T}=T\left( \left| a\right\rangle \left\langle b\right| +\left|
b\right\rangle \left\langle a\right| \right)
\end{equation}
We assume that the tunneling amplitude is much smaller than the
time-independent energy difference $T\ll \varepsilon _{0}$, whereas $V\left(
t\right) $ can be of arbitrary magnitude. The tunneling to and from the
reservoirs brings the system into a mixed state which can only be described
by a density operator $\hat{\rho}$ which obeys the Neumann-Liouville
equation with dissipative terms added to the right-hand-side~\cite
{bib:Nazarov,bib:Gurvitz}: 
\begin{equation}
\partial _{t}\hat{\rho}=-\text{i}\left[ {\cal H},\hat{\rho}\right] +{\cal L}%
_{\text{inc}}\hat{\rho}.  \label{eq:Liou}
\end{equation}
Since the incoherent part in equation (\ref{eq:seq}) and (\ref{eq:cot}) is
invariant under a phase transformation of the nondiagonal elements, we can
derive from (\ref{eq:Liou}) a set of equations that describes the dynamics
on large timescales by first performing a standard {\em time-dependent}
basis transformation~\cite{bib:StoofNazarov} on the density matrix. A
rapidly varying time-dependent phase factor is absorbed in the nondiagonal
elements of $\hat{\rho}$ 
\begin{equation}
\rho _{ab}=\rho _{ab}^{\prime }e^{-i\phi (t)}
\end{equation}
and the diagonal elements are left unchanged. In this new basis the
generalized Liouville equation for $\hat{\rho}$ takes the same form as
equation (\ref{eq:Liou}) with the same incoherent part and a new Hamiltonian 
${\cal H}^{{\cal \prime }}\left( t\right) ={\cal H}_{0}^{{\cal \prime }}+%
{\cal H}_{T}^{{\cal \prime }}\left( t\right) $ with a renormalized energy
difference and a time-dependent tunnel amplitude 
\begin{eqnarray}
{\cal H}_{0}^{\prime } &=&\frac{1}{2}\left( \varepsilon (t)-\partial
_{t}\phi (t)\right) \left( \left| a^{\prime }\right\rangle \left\langle
a^{\prime }\right| -\left| b^{\prime }\right\rangle \left\langle b^{\prime
}\right| \right) , \\
{\cal H}_{T}^{\prime }\left( t\right) &=&Te^{-\text{i}\phi (t)}\left( \left|
a^{\prime }\right\rangle \left\langle b^{\prime }\right| +\left| b^{\prime
}\right\rangle \left\langle a^{\prime }\right| \right) .
\end{eqnarray}
We choose the phase to be $\phi \left( t\right) =n\omega t+(V(t)/\omega
)\sin (\omega t)$ to obtain time-independent diagonal elements 
\begin{equation}
\varepsilon (t)-\partial _{t}\phi (t)=\varepsilon _{0}-n\omega
\end{equation}
which vanish at the $n$-photon resonance $n\omega =\varepsilon _{0}$ in
which we are interested. Furthermore, on small timescales the new tunnel
matrix element is a periodic function of time and can be expanded in a
Fourier series: 
\begin{eqnarray}
{\cal H}_{T}^{\prime }\left( t\right) &\approx &\sum_{m=-\infty }^{+\infty
}e^{-\text{i}\left( m+n\right) \omega t}{\cal H}_{T_{m}}^{\prime }\left(
t\right) , \\
{\cal H}_{T_{m}}^{\prime }\left( t\right) &=&J_{m}\left( 
{\textstyle{V(t) \over \omega }}%
\right) T\left( \left| a^{\prime }\right\rangle \left\langle b^{\prime
}\right| +\left| b^{\prime }\right\rangle \left\langle a^{\prime }\right|
\right) \text{.}
\end{eqnarray}
Likewise we expand the density operator into rapidly oscillating
contributions with amplitudes which vary on large timescales: 
\begin{equation}
\hat{\rho}^{\prime }\left( t\right) =\sum_{r=0}^{\infty }\hat{\rho}^{\prime
\left( r\right) }\left( t\right) e^{\text{i}r\omega t}.
\end{equation}
Inserting this into the generalized Liouville equation for $\hat{\rho}%
^{\prime }$ we obtain an infinite set of coupled equations for the slowly
varying coefficients $\hat{\rho}^{\prime \left( r\right) }$. The amplitude
of the fast oscillations $\hat{\rho}^{\left( 1\right) }$ is of order $%
T/\omega \approx T/\varepsilon _{0}\ll 1$ and can be disregarded: the
``D.C.'' component $\rho ^{\left( 0\right) }\left( t\right) $ satisfies 
\begin{equation}
\partial _{t}\hat{\rho}^{\prime \left( 0\right) }=-\text{i}\left[ {\cal H}%
_{0}^{\prime }+{\cal H}_{T_{-n}}^{\prime },\hat{\rho}^{\prime \left(
0\right) }\right] +{\cal L}_{\text{inc}}\hat{\rho}^{\prime \left( 0\right) }
\end{equation}
Thus the nearly isolated states $|a\rangle $, $|b\rangle $ irradiated at a
resonant frequency $\omega \approx \varepsilon _{0}/n$ are equivalent to
almost degenerate states $|a^{\prime }\rangle $, $|b^{\prime }\rangle $
coupled by a tunneling matrix element 
\begin{equation}
\bar{T}\left( t\right) =J_{-n}\left( 
{\textstyle{V(t) \over \omega }}%
\right) T=\left( -1\right) ^{n}J_{n}\left( 
{\textstyle{V(t) \over \omega }}%
\right) T
\end{equation}
which only varies on large timescales. In the following we will only
consider the $1$-photon resonance i.e. $\omega \approx \varepsilon _{0}$ and
we omit the superscripts used above to distinguish the slowly varying
components from $\hat{\rho}$ itself. However, the equations can be
generalized to the $n$-photon case by replacing $J_{1}\left( V/\omega
\right) \rightarrow J_{n}\left( V/\omega \right) $ and $\varepsilon
_{0}-\omega \rightarrow \varepsilon _{0}-n\omega $. Due to the oscillatory
behavior of the Bessel\ function $J_{1}$ the coherent tunneling amplitude
can be tuned between $0\leq \bar{T}\lesssim $ $0.58T$ by varying the
amplitude/frequency ratio of the irradiation over a range $0\leq V\lesssim
1.84\omega $. The vanishing of the effective tunnel matrix element, $\bar{T}%
=0$\ for non-zero $V/\omega $\ is one of the features which distinguishes
photon-assisted tunneling from adiabatic electron transfer. A similar
renormalization of the tunnel coupling to zero is also known from driven
double well potentials~\cite{bib:GrifoniHanggi}.

The approach developed above allows us the extract the slowly varying
components of the current in both regimes of operation of the electron pump.
We point out that we have only changed the coherent part of the dynamics
which is the same in both regimes.

\subsection{Sequential tunneling regime}

\label{sec:sequential}

The equations (\ref{eq:seq}) describe the dynamics of the double dot in the
sequential tunneling regime on the long timescale when we replace $%
\varepsilon \left( t\right) \rightarrow \varepsilon _{0}-\omega
,T\rightarrow \bar{T}=-J_{1}\left( 
{\textstyle{V(t) \over \omega }}%
\right) T$. The solution of these equations tends to a stationary value,
which is independent of the initial conditions, on typical timescale $\max
\{(\Gamma _{L}^{0,1})^{-1},(\Gamma _{R}^{0,1})^{-1}\}$. The solution of $%
\partial _{t}\hat{\rho}\left( t\right) =0$ gives a Lorentzian lineshape of
the current peak as a function of $\varepsilon _{0}-\omega $ 
\begin{equation}
I/e=I_{\text{max}}w^{2}/[w^{2}+(\varepsilon _{0}-\omega )^{2}],
\label{eq:lorentzian}
\end{equation}
where the maximum current $I_{\text{max}}$\ and half-width at half-maximum $%
w $\ are given by

\begin{eqnarray}
I_{\text{max}}/e &=&1/\left\{ 
{\textstyle{1 \over \Gamma _{L}^{0}+\Gamma _{R}^{1}}}%
\left[ 2+\left( 
{\textstyle{\Gamma _{R}^{0} \over \Gamma _{L}^{0}}}%
+%
{\textstyle{\Gamma _{L}^{1} \over \Gamma _{R}^{1}}}%
\right) \right] +%
{\textstyle{\Gamma _{L}^{0}+\Gamma _{R}^{1} \over 4\bar{T}^{2}}}%
\right\} ,  \label{eq:ImaxSeq} \\
w &=&\sqrt{\left[ 2+\left( 
{\textstyle{\Gamma _{R}^{0} \over \Gamma _{L}^{0}}}%
+%
{\textstyle{\Gamma _{L}^{1} \over \Gamma _{R}^{1}}}%
\right) \right] \bar{T}^{2}+%
{\textstyle{(\Gamma _{L}^{0}+\Gamma _{R}^{1})^{2} \over 4}}%
}
\end{eqnarray}
Let's assume that the tunnel rates for each barrier are the same: $\Gamma
_{L,R}^{0,1}=\Gamma _{L,R}$. If the double dot is weakly coupled to the
leads i.e. $\Gamma _{L,R}\ll T\ll \varepsilon _{0}\approx \omega $ then one
can access the regime $\Gamma _{L,R},\left| \varepsilon _{0}-\omega \right|
\ll \bar{T}$ by adjusting the irradiation amplitude to $V\approx 1.84\omega $%
, where the coherent state in the double dot dominates the transport
properties . The two delocalized states in the double dot are independent
channels for transport and the current increases with $\Gamma _{L,R}$: 
\begin{equation}
I_{\text{max}}/e=1/\left( 
{\textstyle{1 \over \Gamma _{L}}}%
+%
{\textstyle{1 \over \Gamma _{R}}}%
\right) ,\qquad w=2\bar{T}
\end{equation}
In the opposite regime $\left| \varepsilon _{0}-\omega \right| \ll \bar{T}%
\ll \Gamma _{L,R}$, which can be accessed by tuning $V\ll 1.84\omega $, the
decoherence due to tunneling to and from the reservoir dominates the
transport. In this case the height of the current peak is proportional to $%
\bar{T}^{2}$ and {\em decreases} with enhanced tunneling $\Gamma _{L,R}$: 
\begin{equation}
I_{\text{max}}/e=4\bar{T}^{2}/\left( \Gamma _{L}+\Gamma _{R}\right) ,\qquad
w=\left( \Gamma _{L}+\Gamma _{R}\right) /2
\end{equation}
The current peak $I_{\text{max}}/e$ reaches a maximum $\bar{T}/2$ as a
function of $\Gamma _{L,R}$ when $\Gamma _{L}=\Gamma _{R}=2\bar{T}$. This
can be understood as the precise matching of tunneling times: the time for
half a Rabi oscillation in the double dot is exactly equal to the time for
filling the left dot and for emptying the right dot. In Fig. \ref{fig:Iw}(a)
we have plotted the maximum current and the width as a function of the
tunnel rate $\Gamma $ relative to the tunnel coupling $T$. If the double dot
is strongly coupled to the leads i.e. $T\ll \Gamma _{L,R}\ll \varepsilon
_{0}\approx \omega $ only the regime $\bar{T},\left| \varepsilon _{0}-\omega
\right| \ll \Gamma _{L,R}$ is accessible.

We point out that in the sequential tunneling the transport on ``large''
timescales is equivalent to transport of ``free'' electrons (i.e. negligible
interdot repulsion) through a double dot with renormalized static parameters 
$\varepsilon _{0}\rightarrow \varepsilon _{0}-\omega $, $T\rightarrow
J_{1}\left( V/\omega \right) T$, $\Gamma _{L,R}$~\cite{bib:Gurvitz}. Also,
the results here are similar to the analytical results obtained for the
non-interacting case in~\cite{bib:StaffordWingreen}, where only sequential
tunneling.

\subsection{Co-tunneling regime}

In the co-tunneling regime two well-separated timescales are involved,
namely the ``long'' time $t_{\text{ct}}=\Gamma _{\text{ct}}^{-1}$ between
two co-tunneling processes and the ``short'' time of the process itself $t_{%
\text{virt}}=\varepsilon _{0}^{-1}$ during which the energy is uncertain by
an amount $\varepsilon _{0}$. Near resonance the applied frequency matches
the detuning of the levels so $t_{\text{virt}}\approx 2\pi /\omega
=t_{\omega }$. The density matrix approach developed above describes the
system on the ``large'' timescale $t_{\text{ct}}$. The equations (\ref
{eq:cot}) describe the dynamics of the double dot in the sequential
tunneling regime on the long timescale when we replace $\varepsilon \left(
t\right) \rightarrow \varepsilon _{0}-\omega ,T\rightarrow \bar{T}%
=-J_{1}\left( 
{\textstyle{V(t) \over \omega }}%
\right) T.$ The stationary solution of these equations gives a Lorentzian
current in $\varepsilon _{0}-\omega $ with height $I_{\text{max}}$ and
half-width $w$: 
\begin{eqnarray}
I_{\text{max}}/e &=&1/\left( 
{\textstyle{2 \over \Gamma _{\text{ct}}}}%
+%
{\textstyle{\Gamma _{\text{ct}} \over 4\bar{T}^{2}}}%
\right) ,  \label{eq:ImaxCot} \\
w &=&\sqrt{2\bar{T}^{2}+%
{\textstyle{\Gamma _{\text{ct}}^{2} \over 4}}%
}.
\end{eqnarray}
For weak coupling to the reservoirs $\Gamma _{\text{ct}}\ll \bar{T}$ the
peak height is constant and the width increases linearly with the coherent
coupling $\bar{T}$: 
\[
I_{\text{max}}/e=\frac{1}{2}\Gamma _{\text{ct}},\qquad w=\sqrt{2}\bar{T} 
\]
In the opposite case $\Gamma _{\text{ct}}\gg \bar{T}$ the peak height is
small but increases rapidly with $\bar{T}$ whereas now the width is
constant: 
\[
I_{\text{max}}/e=4\bar{T}^{2}/\Gamma _{\text{ct}},\qquad w=\frac{1}{2}\Gamma
_{\text{ct}} 
\]
The current peak $I_{\text{max}}/e$ reaches a maximum $\bar{T}/\sqrt{2}$ as
a function of $\Gamma _{\text{ct}}$ when $\Gamma _{\text{ct}}=2\sqrt{2}\bar{T%
}$. In Fig. \ref{fig:Iw}(b) the scaled pumping current is plotted for
different values of $\Gamma _{\text{ct}}/\bar{T}$. In the co-tunneling
regime electrons are transferred through strongly correlated transport
channels. Therefore the condition for the maximum current cannot be
understood as a the precise matching of tunneling times as in the sequential
tunneling regime (factor $\sqrt{2})$. Equations (\ref{eq:cot}) coincide with
those that describe the transport of electrons which are correlated by
strong Coulomb repulsion through a double dot\ at high voltage bias, with
renormalized static parameters $\varepsilon _{0}\rightarrow \varepsilon
_{0}-\omega $, $T\rightarrow J_{1}\left( V/\omega \right) T$, $\Gamma _{%
\text{ct}}\rightarrow \Gamma _{R}$ in the limit where the tunneling to the
into the left dot is so fast, $\Gamma _{L}\gg \Gamma _{R},T$, that the
current doesn't depend on it anymore\cite{bib:StoofNazarov}. In this system
the correlation of the conduction channels prevents more than one channel
from being occupied and reduces the effective tunnel rate by a factor $2$ in
(\ref{eq:ImaxCot})\cite{bib:WegewijsNazarov}. This blockade can be obtained
formally from the density matrix equations of the ``free'' electrons (\ref
{eq:seq}) by taking $\Gamma _{L,R}^{1}\rightarrow 0$ and $\Gamma _{L}^{0}\gg
\Gamma _{R}^{0}=\Gamma _{\text{ct}}$.

Comparing the co-tunneling and sequential tunneling regime the main
difference is that for fixed $\Gamma _{L,R}$\ the co-tunnel rate $\propto
\Gamma _{L}\Gamma _{R}$\ is much smaller: $\Gamma _{\text{ct}}\ll \Gamma
_{L,R}$. For $\Gamma _{L,R}\sim \bar{T}$\ the sequential tunneling current
can be near its maximal value $\sim 2\bar{T}$\ , whereas the co-tunneling
current will be $\propto \Gamma _{\text{ct}}\ll \bar{T}$. However, for $%
\Gamma _{L,R}\gg \bar{T}$\ the sequential tunneling current is much smaller
than $\bar{T}$\ and it is possible to adjust $\Gamma _{\text{ct}}\sim \bar{T}
$\ so that the co-tunneling current takes its maximal value $\sim 2\sqrt{2}%
\bar{T}$\ which is much larger. Thus for fixed $\bar{T}$\ the maximal
co-tunneling current is larger by a factor $\sqrt{2}$\ than the maximal
sequential tunneling current (Fig. \ref{fig:Iw}). The width of the co-tunnel
peak is also smaller than in the sequential tunneling regime.

\section{Modulated irradiation of a double dot}

\label{sec:modulation}In this section we consider the sequential tunneling
regime already discussed in section \ref{sec:sequential} and apply our
approach for large timescales to the cases of pulsed and slow sinusoidal
modulation of the irradiation amplitude.

\subsection{Response to irradiation pulses}

By means of irradiation pulses quantum states in the dots can be
manipulated. Assuming the double dot to be in the ground state 
\begin{equation}
\rho _{00,00}=1,\rho _{-10,-10}=\rho _{01,01}=\rho _{-11,-11}=\rho
_{-11,00}=\rho _{00,-11}=0  \label{eq:rhoinitial}
\end{equation}
at $t=0$ we solve equations (\ref{eq:seq})\ with $\varepsilon \left(
t\right) \rightarrow \varepsilon _{0}-\omega ,T\rightarrow \bar{T}%
=-J_{1}\left( 
{\textstyle{V(t) \over \omega }}%
\right) T$ for the time-evolution under the influence of the irradiation for
the case $\Gamma _{L,R}^{0,1}=\Gamma $: 
\begin{equation}
I\left( t\right) =I\left[ 1-e^{-\Gamma t}\left( \cos \left( \Omega _{\text{R}%
}t\right) +\frac{\Gamma }{\Omega _{\text{R}}}\sin \left( \Omega _{\text{R}%
}t\right) \right) \right]
\end{equation}
where $\Omega _{\text{R}}=\sqrt{\left( \varepsilon _{0}-\omega \right) ^{2}+4%
\bar{T}^{2}}$ is the Rabi frequency. For $t\gg \Gamma ^{-1}$ the solution
tends to the stationary current (\ref{eq:ImaxSeq}) derived before: 
\begin{equation}
I/e=1/\left( 
{\textstyle{1 \over \Gamma }}%
{\textstyle{\Omega _{\text{R}}^{2} \over 2\bar{T}^{2}}}%
+%
{\textstyle{\Gamma  \over 2\bar{T}^{2}}}%
\right)
\end{equation}
When the irradiation is switched off at $t=\tau _{p}$, the current decays
exponentially as $e^{-\Gamma \left( t-\tau _{p}\right) }$ to zero, as can
also be seen by solving the equations for the case $\omega =\bar{T}=0$. One
can resolve the Rabi oscillations in the D.C.-current by considering the
current averaged over a series of identical pulses~\cite{bib:Nakamura} with
delay $\tau _{d}$, as a function of the pulse-length $\tau _{p}$ (see inset
of Fig. \ref{fig:Iperiod}): 
\begin{equation}
I_{\text{dc}}\left( \tau _{p}\right) =\frac{1}{\tau _{d}}\left(
\int_{0}^{\tau _{p}}I\left( t\right) dt+I\left( \tau _{p}\right) \int_{\tau
_{p}}^{\tau _{d}}e^{-\Gamma \left( t-\tau _{p}\right) }dt\right)
\end{equation}
Here $\tau _{d}-\tau _{p}\gg \Gamma ^{-1}$ to ensure that the system is
prepared in the ground state (\ref{eq:rhoinitial}) at the beginning of the
pulse. In the physically interesting case of weak coupling to the leads $%
\Gamma \ll \Omega _{\text{R}}$ we obtain 
\begin{equation}
I_{\text{dc}}\left( \tau _{p}\right) \approx I\frac{1}{\tau _{d}}\left[ \tau
_{p}+\frac{1-e^{-\Gamma \tau _{p}}\cos \left( \Omega _{\text{R}}\tau
_{p}\right) }{\Gamma }\right]
\end{equation}
The oscillations are most clearly resolvable for $\tau _{p}\ll \Gamma ^{-1}$
(Fig. \ref{fig:Iperiod}). The period of the coherent oscillation at
resonance $\omega =\varepsilon _{0}$, $2\pi /\Omega _{\text{R}}=\pi /\left(
J_{1}\left( V_{0}/\omega \right) T\right) $, can be tuned by varying the
irradiation power $V_{0}$.

\subsection{Sinusoidal amplitude modulation}

Now consider the case where the amplitude of the irradiation is slowly
sinusoidally modulated with small modulation amplitude $\tilde{V}\ll V_{0}$: 
\begin{equation}
V(t)=V_{0}+\tilde{V}\cos \left( \Omega t\right) .
\end{equation}
The case where the modulation amplitude is of the order of or larger than
the irradiation amplitude is physically not very interesting because the
system then exhibits trivial Fourier peaks at $\omega $, $\omega +\Omega $
and $\omega -\Omega $ as a function of $\varepsilon _{0}$ in the
D.C.-current. To find an analytical solution, we rewrite equations (\ref
{eq:seq})\ with $\varepsilon \left( t\right) \rightarrow \varepsilon
_{0}-\omega ,T\rightarrow \bar{T}=-J_{1}\left( 
{\textstyle{V(t) \over \omega }}%
\right) T$ in matrix notation: 
\begin{equation}
\frac{\partial \vec{\rho}}{\partial t}=\left( \hat{\Gamma}+\hat{T}\right) 
\vec{\rho}+\vec{c},
\end{equation}
where $\vec{\rho}=\left( \rho _{-10,-10,}\rho _{00,00},\rho _{-11,-11},\rho
_{01,01},\rho _{-11,00}\right) ^{T}$, $\vec{c}=\left( \Gamma
_{R}^{1},0,0,0,0\right) ^{T}$and expand the Bessel function $J_{1}\left(
V\left( t\right) /\omega \right) $ in a Taylor series to second order in $%
\tilde{V}$. If we now consider the Fourier coefficients $\vec{\rho}_{n}$\ of 
$\vec{\rho}\left( t\right) $\ and $T_{n}$\ of $J_{1}\left( V\left( t\right)
/\omega \right) T$, we find the following equations for the D.C.-component
and first harmonic, if we disregard higher harmonics 
\begin{eqnarray}
0 &=&\left( \hat{\Gamma}+T_{0}\hat{T}\right) \vec{\rho}_{0}+T_{1}\hat{T}%
\left( \vec{\rho}_{+1}+\vec{\rho}_{-1}\right) +\vec{c}, \\
\vec{\rho}_{\pm 1} &=&-T_{1}\left( \hat{\Gamma}+T_{0}\hat{T}\mp i\Omega \hat{%
I}\right) ^{-1}\hat{T}\vec{\rho}_{0},
\end{eqnarray}
from which we can solve for the D.C.-component $\vec{\rho}_{0}.$ If we
furthermore assume that $\Gamma _{L,R}^{0,1}=\Gamma \ll \bar{T}$, the
solution shows additional sidepeaks in the photoresponse of the system,
which are in good approximation Lorentzians along the hyperbola $\varepsilon
_{0}-\omega =\sqrt{\Omega ^{2}-4\bar{T}^{2}}$ i.e. $\Omega =\Omega _{\text{R}%
}$ as plotted in Fig. (\ref{fig:sinusmod}). The height of these peaks is 
\begin{equation}
\frac{I_{\text{max}}}{e}=\frac{\Gamma }{2}%
{\displaystyle{\left( \alpha ^{2}-4\right) ^{2}\bar{T}^{^{\prime }2}\frac{\tilde{V}^{2}}{\omega ^{2}} \over \alpha ^{2}\left( \alpha ^{2}\Gamma ^{2}+\left( \alpha ^{2}-4\right) \bar{T}^{^{\prime }2}\frac{\tilde{V}^{2}}{\omega ^{2}}\right) }}%
,
\end{equation}
with $\alpha =\Omega /\bar{T},\bar{T}^{^{\prime }}=J_{1}^{^{\prime
}}(V_{0}/\omega )T$ and the half-width at half maximum is 
\begin{equation}
w=\sqrt{\bar{T}^{^{\prime }2}\frac{\tilde{V}^{2}}{\omega ^{2}}+\frac{\alpha
^{2}}{\alpha ^{2}-4}\Gamma ^{2}}
\end{equation}
These sidepeaks thus resolve the Rabi splitting as described in Refs.~\cite
{bib:StaffordWingreen,bib:Holthaus} in terms of quasi-energies. The height
of these peaks can be of the order of the first satellite peak and should
therefore be experimentally observable. Notice that at $\varepsilon
_{0}=\omega ,\Omega =2\bar{T}$ the width diverges; however, the current peak
at this point vanishes since the exact matching $\varepsilon _{0}=\omega $
gives a\ resonance to which the modulation cannot add extra current.

Thus by applying a high frequency $\omega $ we reduce the energy spacing by
a large amount, $\varepsilon _{0}\rightarrow \varepsilon _{0}-\omega $, and
modify the tunneling matrix element with an intensity ($V$) dependent
factor, $T\rightarrow \bar{T}$. The lower frequency $\Omega $ allows one to
precisely match the remaining small energy difference $\varepsilon
_{0}-\omega $ without significantly altering the tunneling matrix element $%
\bar{T}$, thereby inducing a photo-current.

\section{Conclusions}

\label{sec:conclusion}We have considered an{\bf \ }electron pump consisting
of an double quantum dot subject to irradiation. An incoherent and a new
coherent pumping mechanism were discussed. We have derived equations of
motion for the density matrix elements of the double-dot system which are\
time-averaged over an interval which is long compared to the period of the
applied signal. From these equations we calculated the pumping current in
both regimes. In both cases the current peak is a Lorentzian. Surprisingly,
for fixed effective tunnel coupling the maximal pumping current in the
co-tunneling regime is larger by a factor $\sqrt{2}$ compared to the value
in the sequential tunneling regime, where the maximum occurs at a different
value of the tunnel rates for each regime. Experimental realization of this
device would allow for a systematic study of coherent transport through a
solid-state qubit. Moreover, modulation of the irradiation amplitude
exhibits interesting phenomena: a train of short pulses should resolve the
Rabi frequency in the time-averaged current as a function of the pulse
length, and sinusoidal amplitude modulation should provide a tool to resolve
the Rabi splitting.

It is a pleasure to acknowledge useful discussions with Gerrit Bauer, Arne
Brataas, Leo Kouwenhoven and Wilfred van der Wiel. This work is part of the
research program of the ``Stichting voor Fundamenteel Onderzoek der
Materie'' (FOM), which is financially supported by the ''Nederlandse
Organisatie voor Wetenschappelijk Onderzoek'' (NWO) and the NEDO project
NTDP-98.

\begin{figure}[tbp]
\caption{Double dot electron pump. The asymmetric gate voltages allow one to
induce a photo-current at zero bias. One can excite an electron from the
highest level filled in the left dot to the first empty level of the right
dot with an AC field that is resonant with this transition, but not with the
corresponding transition in the opposite direction.}
\label{fig:2dotspump}
\end{figure}

\begin{figure}[tbp]
\caption{Stability diagram of the double dot system near the region where
the state $|0,0\rangle $ is stable, with $N_{1}$ and $N_{2}$ electrons in
dot $1$ and $2$, respectively. Along the horizontal and vertical axis the
sequential tunneling barrier $\Delta _{L}^{-}$ and $\Delta _{R}^{+},$%
respectively, are varied independently.}
\label{fig:stab}
\end{figure}

\begin{figure}[tbp]
\caption{{}Energy diagrams of the double dot electron pump together with the
leads: a) sequential tunneling regime, b) cotunneling regime. The Coulomb
interaction between the extra positive charge on the left dot and negative
charge on the right dot determines the position of level $E_{-1,1}$ relative
to $E_{-1,0}$ and $E_{0,1}$.}
\label{fig:energies}
\end{figure}

\begin{figure}[tbp]
\caption{Lorentzian current peak height (bold solid line) and width (dashed
line) as a function of a) $\Gamma /\bar{T}$ ($\Gamma =\Gamma _{L}=\Gamma _{R}
$) in the sequential tunneling regime and of b) $\Gamma _{\text{ct}}/\bar{T}$
in the cotunneling regime. For fixed $\bar{T}$\ the maximal co-tunneling
current as a function of $\Gamma _{\text{ct}}$ is larger than the maximal
sequential tunneling current as a function of $\Gamma $.}
\label{fig:Iw}
\end{figure}

\begin{figure}[tbp]
\caption{Double dot subjected to an irradiation pulse train. Figure:{\em \
Current averaged over a series of identical pulses},in units of number of
electrons transferred per pulse, as a function of the pulse length for $%
\protect\varepsilon _{0}=$ $\protect\omega =30$, $T=1$, $\Gamma =0.1$ and
several irradiation amplitudes\thinspace $V$. Inset: {\em time dependent
current} during one pulse.}
\label{fig:Iperiod}
\end{figure}

\begin{figure}[tbp]
\caption{Electron pump with harmonically amplitude-modulated irradiation.
Current as a function of the renormalized level detuning (resonance mismatch
of basis frequency) and the frequency of the amplitude modulation for $T=1$, 
$\Gamma _{\text{L,R}}=0.2$, $V_{0}=30$, $\tilde{V}=10$.}
\label{fig:sinusmod}
\end{figure}

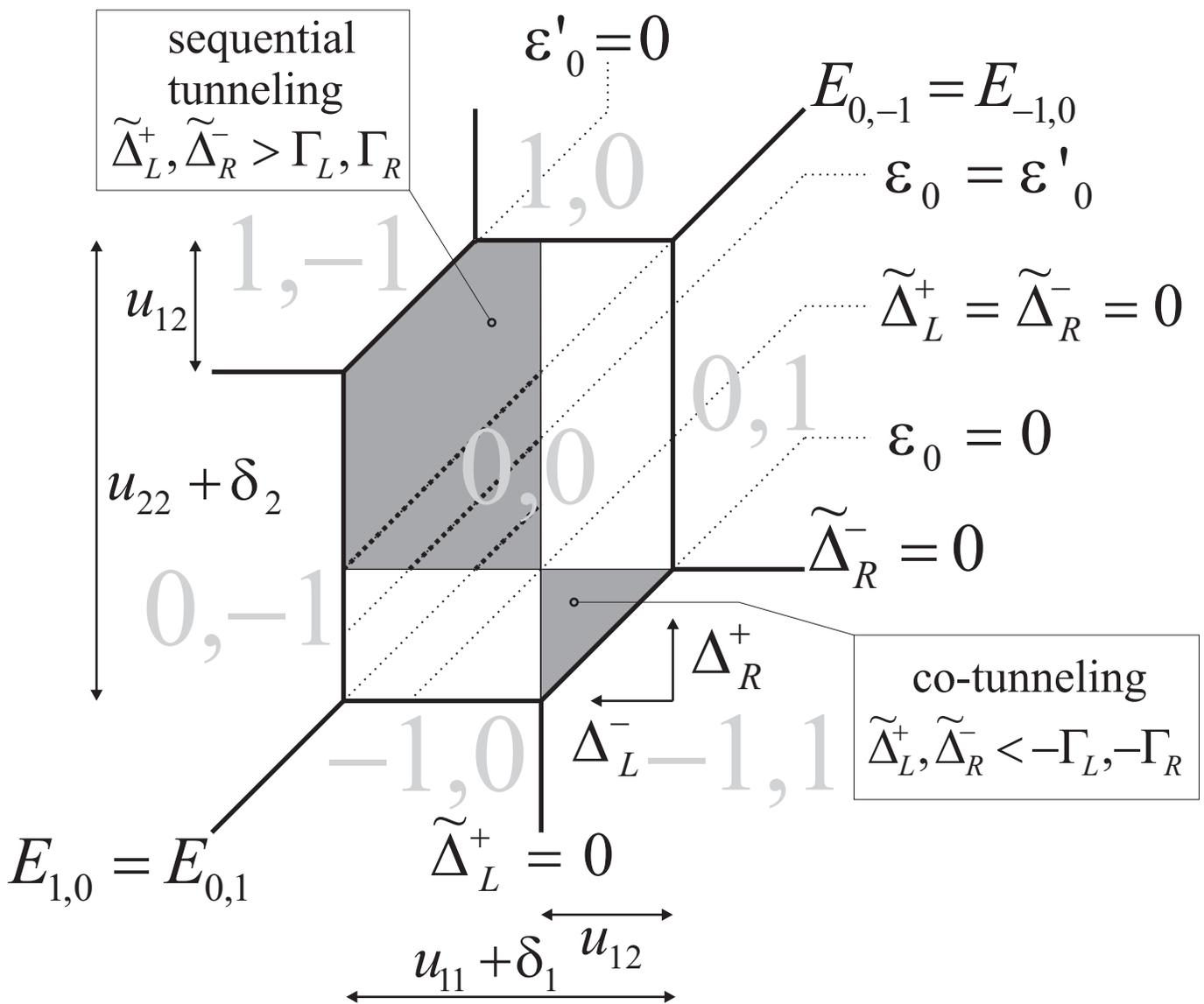
\end{document}